\documentclass[final,5p,times,twocolumn,authoryear]{elsarticle}

\makeatletter
\def\ps@pprintTitle{%
 \let\@oddhead\@empty
 \let\@evenhead\@empty
 \def\@oddfoot{\footnotesize\itshape Published in \href{https://doi.org/10.1016/j.epsl.2018.11.034}{Earth and Planetary Science Letters}\hfill\today}%
 \let\@evenfoot\@oddfoot}
\makeatother

\usepackage{amssymb}

\usepackage{lineno}

\usepackage{doi}
\usepackage{xspace}
\usepackage{nameref}
\usepackage{natbib}
\usepackage{enumerate}
\usepackage[fleqn]{amsmath}
\usepackage{hyperref} 
\usepackage{color}
\usepackage{soul}
\usepackage[normalem]{ulem}
\usepackage{threeparttable}
\usepackage{placeins}

\usepackage[scaled]{helvet}
 
\usepackage[T1]{fontenc}

\newcommand\pnas{{Proc. Natl. Acad. Sci. }}
\newcommand\apj{{Astrophys. J. }}%
\newcommand\aap{{Astron. Astrophys. }}%
\newcommand\epsl{{Earth Planet. Sci. Lett. }}%
\newcommand\icarus{{Icarus }}%
\newcommand\mnras{{Mon. Not. R. Astron. Soc. }}%
\newcommand\ssr{{Space~Sci.~Rev. }}%
\newcommand\nat{{Nature }}%
\newcommand\gca{{Geochim.~Cosmochim.~Acta }}%
\newcommand\jgrp{{J.~Geophys.~Res.~Planets }}%
\newcommand\planss{{Planet.~Space~Sci. }}%
\usepackage[dvipscolors]{xcolor}
\hypersetup{ 
    colorlinks=true,
    linkcolor=blue,
    filecolor=blue,      
    urlcolor=blue,
    citecolor=blue,
}
\definecolor{tkcl}{RGB}{251,180,174} 
\definecolor{rfkcl}{RGB}{179,205,227} 
\definecolor{gjgcl}{RGB}{204,235,197} 
\definecolor{tvgcl}{RGB}{222,203,228} 
\definecolor{tlcl}{RGB}{254,217,166} 
\definecolor{mrmcl}{RGB}{255,255,204} 
\definecolor{redhl}{rgb}{1,0.5,.5} 
\definecolor{bluehl}{rgb}{0.75,0.75,1} 
\definecolor{greenhl}{rgb}{0.5,1.0,0.5} 
\definecolor{yellowhl}{rgb}{1.0,1.0,0.5} 
\definecolor{brightgreen}{RGB}{140,235,174} 
\definecolor{brightblue}{RGB}{140,202,235}

\begin{document}

\begin{frontmatter}



\title{Magma ascent in planetesimals: control by grain size}


\author[inst1]{Tim Lichtenberg\corref{cor1}}
\author[inst2]{Tobias Keller}
\author[inst3]{Richard F. Katz}
\author[inst4]{Gregor J. Golabek}
\author[inst1]{Taras V. Gerya}

\address[inst1]{Institute of Geophysics, ETH Z{\"u}rich, Sonneggstrasse 5, 8092 Z{\"u}rich, Switzerland}
\address[inst2]{Department of Geophysics, Stanford University, Stanford, CA 94305, United States}
\address[inst3]{Department of Earth Sciences, University of Oxford, Oxford OX1 3AN, United Kingdom}
\address[inst4]{Bayerisches Geoinstitut, University of Bayreuth, Universit{\"a}tsstrasse 30, 95440 Bayreuth, Germany}

\cortext[cor1]{Corresponding author, now at: Atmospheric, Oceanic and Planetary Physics, University of Oxford, Parks Rd, Oxford OX1 3PU, United Kingdom, email: \href{mailto:tim.lichtenberg@physics.ox.ac.uk}{tim.lichtenberg@physics.ox.ac.uk}.}

\begin{abstract}

Rocky planetesimals in the early solar system melted internally and evolved chemically due to radiogenic heating from $^{26}$Al. Here we quantify the parametric controls on magma genesis and transport using a coupled petrological and fluid mechanical model of reactive two-phase flow. We find the mean grain size of silicate minerals to be a key control on magma ascent. For grain sizes $\gtrsim$ 1 mm, melt segregation produces distinct radial structure and chemical stratification. This stratification is most pronounced for bodies formed at around 1 Myr after formation of Ca,Al-rich inclusions. These findings suggest a link between the time and orbital location of planetesimal formation and their subsequent structural and chemical evolution. According to our models, the evolution of partially molten planetesimal interiors falls into two categories. In the \emph{magma ocean} scenario, the whole interior of a planetesimal experiences nearly complete melting, which would result in turbulent convection and core-mantle differentiation by the rainfall mechanism. In the \emph{magma sill} scenario, segregating melts gradually deplete the deep interior of the radiogenic heat source. In this case, magma may form melt-rich layers beneath a cool and stable lid, while core formation would proceed by percolation. Our findings suggest that grain sizes prevalent during the internal heating stage governed magma ascent in planetesimals. Regardless of whether evolution progresses toward a \emph{magma ocean} or \emph{magma sill} structure, our models predict that temperature inversions due to rapid $^{26}$Al redistribution are limited to bodies formed earlier than $\approx$ 1 Myr after CAIs. We find that if grain size was $\lesssim$ 1 mm during peak internal melting, only elevated solid--melt density contrasts (such as found for the reducing conditions in enstatite chondrite compositions) would allow substantial melt segregation to occur. \newline

\end{abstract}


\end{frontmatter}

\section{Introduction}
\label{sec:introduction}

At the time of planet formation, the inner solar system was populated by rocky planetesimals that seeded today's terrestrial planets through dynamical accretion of many smaller bodies \citep{2015SciA....115109J}, and whose internal evolution was governed by radiogenic heating from short-lived $^{26}$Al \citep{2006M&PS...41...95H}. For large planetesimal radii and sufficient $^{26}$Al incorporated upon formation, the released energy led to volatile degassing \citep{castillorogez2017,Monteux2018} and significant silicate melting, surpassing the rheological transition from solid-state creep to disaggregation and melt-dominated deformation at melt fractions $\phi_{\mathrm{trans}}$ $\gtrsim 0.4$--$0.6$ \citep[][]{2009GGG....10.3010C}. In comparison with solid or partially molten interiors, which lose heat by conduction and/or laminar convection, disaggregation results in significantly increased heat flux by turbulent convection and rapid metal-silicate differentiation by the raining out of iron droplets \citep{1990orea.book..231S}.

The interior evolution of early solar system planetesimals has broad implications for the formation of both rocky planets and main-belt asteroid populations, the most immediate remnants of the accretion process. Meteorites, broken-up pieces of asteroids fallen to Earth, are currently our only source of direct evidence from the early solar system. Therefore, our understanding of planetary growth and evolution is fundamentally limited by our ability to reconstruct the thermo-chemical evolution of planetesimals as evidenced by meteorites. Achondritic meteorites, which are thought to originate from differentiated planetesimals, show a remarkable diversity and likely originate from more than 50--100 parent bodies \citep{1990Sci...249..900W}. However, spectral properties of asteroids do not match this diversity, as most known asteroids with an achondritic surface are interpreted to represent the debris of only a few parent bodies \citep{burbine2017}. This apparent lack of achondritic asteroids is at odds with the available meteorite record.

A possible solution to this conundrum is that internally differentiated planetesimals can retain their primitive, chondritic surfaces if magma remains confined to the interior instead of being erupted by volcanism \citep{2011E&PSL.305....1E,2013AREPS..41..529W}. Some paleomagnetic studies on CV and CM meteorites suggest previous dynamo activity consistent with this hypothesis \citep{2011PNAS..108.6386C,2015E&PSL.410...62C}. Furthermore, Rosetta spacecraft data indicates a carbonaceous or enstatite chondrite surface for 21 Lutetia and an average density of $\approx$ 3400 kg m$^{-3}$ \citep{2011Sci...334..487S_short,2011Sci...334..491P}, higher than known chondrites and consistent with past compaction and partial melting beneath a primitive, chondritic crust \citep{2012P&SS...66..137W,2013Icar..224..126N}.

Based on the available evidence, most current models propose a magma ocean scenario, where high melt fractions ($\phi \gtrsim$ $\phi_{\mathrm{trans}}$) dominated the thermal and chemical evolution of planetesimal interiors. For the purposes of this study, we characterize the \emph{magma ocean} scenario as a planetesimal exhibiting a fully molten interior of a well-mixed composition and an adiabatic temperature profile located beneath a thin ($\approx$ 10 km), unmolten, chemically primitive lid with a linear conductive thermal profile. Recent modeling studies investigating this scenario have relied either on thermal modeling with parameterized melting \citep[e.g.,][]{2006M&PS...41...95H,2011E&PSL.305....1E}, or on one-phase convection models \citep[e.g.,][]{2014MPS...49.1083G,2016Icar..274..350L,2018Icar..302...27L} that capture the collective flow and thermo-chemical evolution of partially molten rock or partly crystalline magma.

However, two-phase theory of partially molten systems \citep[e.g.,][]{1984JPet....5..713M} suggests that silicate melts may buoyantly ascend relative to the ambient rock matrix. Depending on the compositional and rheological properties of silicate minerals and their melts, this segregation may have delayed, or even precluded, the generation of a \emph{magma ocean} structure. Ascending melts may instead have formed melt-rich layers beneath the primitive lid, hereafter referred to as \emph{magma sills} \citep{wilson_keil_2017}. For the purposes of this study, we define this \emph{magma sill} scenario as a body with radial heterogeneities of melt fraction that differ from the fiducial \emph{magma ocean} case. This scenario implies a potentially significant redistribution of $^{26}$Al, which is a moderately incompatible element and preferentially partitions into silicate melts. The transfer of the major heat source into shallow \emph{magma sills} might then result in a transient, inverted temperature profile with a thermal history distinct from the \emph{magma ocean} scenario. To date, only few studies have quantitatively investigated this effect. These were either based on melt transport models with parameterized melt ascent velocities \citep{2011MaPS...46..903M,2012ChEG...72..289W,2013M&PS...48.2333M,2013Icar..224..126N,2014E&PSL.395..267N,Neumann2018}, or focused on metal-silicate separation \citep{2012Icar..217..339S,2017PNAS..11413406G}.

The efficiency of melt transport in planetesimals depends on various parameters. The presence of primordial volatiles favors rapid segregation by increasing the buoyancy and lowering the viscosity of magmas. However, if volatiles are exsolved before the onset of silicate melting, \citet{2014E&PSL.390..128F} argue that the segregation rate of dry melt is mostly controlled by the oxygen fugacity and the degree of melting. The oxygen fugacity determines (or is determined by) the relative abundance of FeO and Fe-FeS in the primordial rock, with parts of the latter potentially lost to the core by percolation before the onset of major silicate melting \citep{2003Natur.422..154Y,2009PEPI..177..139B,2015E&PSL.417...67C,2017PNAS..11413406G}. Higher oxygen fugacity may therefore result in more Fe-rich silicate melts with reduced (or even inverted) density contrast relative to the host rock. Lower oxygen fugacity, in contrast, may produce iron-poor, buoyant melts that ascend rapidly.

In this study, we seek to establish regime boundaries that separate primary evolution scenarios of early solar system planetesimals by assessing the effects of melt segregation on thermal evolution and chemical differentiation. We focus on the melting and transport of the major lithophile phases in primitive bodies and investigate the potential for melt accumulation and heat source redistribution. We employ a computational model of coupled fluid dynamics and thermo-chemical evolution that combines multi-component petrological reactions with a two-phase magma transport model. We quantify the leading controls on melt segregation in planetesimals using theoretical considerations and numerical calculations of idealized planetesimal evolution. Our results show that both the \emph{magma ocean} and \emph{magma sill} scenarios are realized within a relevant parameter space. We will focus our discussion on the latter case, where melt segregation is most important.


\section{Melt segregation scaling}
\label{sec:scaling}

To gain a leading-order understanding of silicate melt ascent in $^{26}$Al-heated planetesimals, we first consider the characteristic time scales of melt transport in partially molten bodies. In two-phase theory \citep[][]{1984JPet....5..713M}, the characteristic length scale of melt migration by porous flow relative to a permeable rock matrix is given by the compaction length
\begin{linenomath*}
\begin{align}
    \delta_{\mathrm{c}} = \sqrt{\frac{k_{0} \eta_0}{\phi_0 \mu_0}}, \label{eq:compaction_length}
\end{align}
\end{linenomath*}
with characteristic rock viscosity $\eta_0$, melt viscosity $\mu_0$, melt fraction $\phi_0$, and rock permeability
\begin{linenomath*}
\begin{align}
    k_{\mathrm{0}} = \frac{a_0^2}{b} \frac{\phi_0^{\mathrm{n}}}{(1-\phi_0)^{\mathrm{m}}} \: , \label{eq:permeability}
\end{align}
\end{linenomath*}
where $a_0$ is the characteristic grain size, $b$ a geometric factor, and $m$, $n$ powerlaw exponents. The characteristic velocity of segregating melts is
\begin{linenomath*}
\begin{align}
    w_0 = \frac{k_{\mathrm{0}} \Delta \rho_0 g_0}{\mu_0 \phi_0} \: , \label{eq:migration_speed}
\end{align}
\end{linenomath*}
with $\Delta \rho_0$ the solid-melt density contrast, and $g_0$ the surface gravity. In primordial, homogeneous planetesimals, gravity increases with increasing distance $r$ from the center,
\begin{linenomath*}
\begin{align} g(r) = g_0r / R_\mathrm{P},
\end{align}
\end{linenomath*}
where $R_\mathrm{P}$ is the total planetesimal radius. The first silicate melts in sufficiently large planetesimals form in an adiabatic zone stretching from the center, where gravity is negligible, to below the upper conductive lid \citep{2006M&PS...41...95H,2016Icar..274..350L}. As melting progresses, the permeability increases and melts in shallower regions of the planetesimal, where gravity is highest, begin to segregate from the residual rock. Melt segregation can alter the chemical and thermal structure of $^{26}$Al-heated planetesimals, because early-formed melts are preferentially enriched in incompatible elements including $^{26}$Al. 

We define the characteristic {\em melt segregation time scale}, $\tau_{\mathrm{segr}}$, as
\begin{linenomath*}
\begin{align}
	\tau_{\mathrm{segr}} = R_{\mathrm{P}} / w_0 \: .
\end{align}
\end{linenomath*}
To achieve a substantial redistribution of heat-producing $^{26}$Al, the rate of melt transport must exceed the rate of melt generation. We thus define the {\em heating time scale}, $\tau_{\mathrm{heat}}$, of a planetesimal at a given time $t_0$ after Ca,Al-rich inclusions (CAIs) as
\begin{linenomath*}
\begin{align}
	\tau_{\mathrm{heat}} = c_{\mathrm{p}} \Delta T_{0} / H_{^{26}\mathrm{Al}}(t_0) \: ,
\end{align}	
\end{linenomath*}
with the specific heat capacity of silicates, $c_{\mathrm{p}} = 1100$ J kg$^{-1}$ K$^{-1}$, the temperature difference between accretion and solidus temperature, $\Delta T_{0} \approx 1100$ K, and the decay power of $^{26}$Al per unit mass,
\begin{linenomath*}
\begin{align}
	H_{^{26}\mathrm{Al}}(t_0) = f_{\mathrm{Al}} \cdot \left[ \frac{^{26}\mathrm{Al}}{^{27}\mathrm{Al}} \right]_0 \cdot \frac{E_{\mathrm{^{26}\mathrm{Al}}}}{\tau_{^{26}\mathrm{Al}}} \cdot \mathrm{e}^{-t_0/\tau_{^{26}\mathrm{Al}}} \: . \label{eq:aldecay}
\end{align}
\end{linenomath*}
Here, $f_{\mathrm{Al}}$ is the chondritic abundance of aluminum per unit mass \citep{2003ApJ...591.1220L}, $\left[ \frac{^{26}\mathrm{Al}}{^{27}\mathrm{Al}} \right]_0 = 5.25 \times 10^{-5}$ is the canonical ratio of $^{26}$Al to $^{27}$Al at CAI formation \citep{2013M&PS...48.1383K}, $E_{\mathrm{^{26}\mathrm{Al}}} = 3.12$ MeV $= 5 \times 10^{-13}$ J is the decay energy, and $\tau_{^{26}\mathrm{Al}} = t_{1/2,^{26}\mathrm{Al}}/\mathrm{ln}(2) = 1.03$ Myr is the mean lifetime of $^{26}$Al.

Using these characteristic scales, we define the non-dimensional {\em melt segregation number}, 
\begin{linenomath*}
\begin{align}
	R_{\mathrm{seg}} = \mathrm{log}_{10} \left( \dfrac{\tau_{\mathrm{heat}} }{ \tau_{\mathrm{segr}}} \right) = \mathrm{log}_{10} \left( \dfrac{ k_{0} \Delta \rho_0 g_0 c_{\mathrm{p}} \Delta T_{0} \phi_0}{ R_{\mathrm{P}} \mu_0 H_{^{26}\mathrm{Al}}(t_0) } \right) \: ,
\end{align}
\end{linenomath*}
\begin{figure}[tbh]
    \centering
    \includegraphics[width=0.45\textwidth]{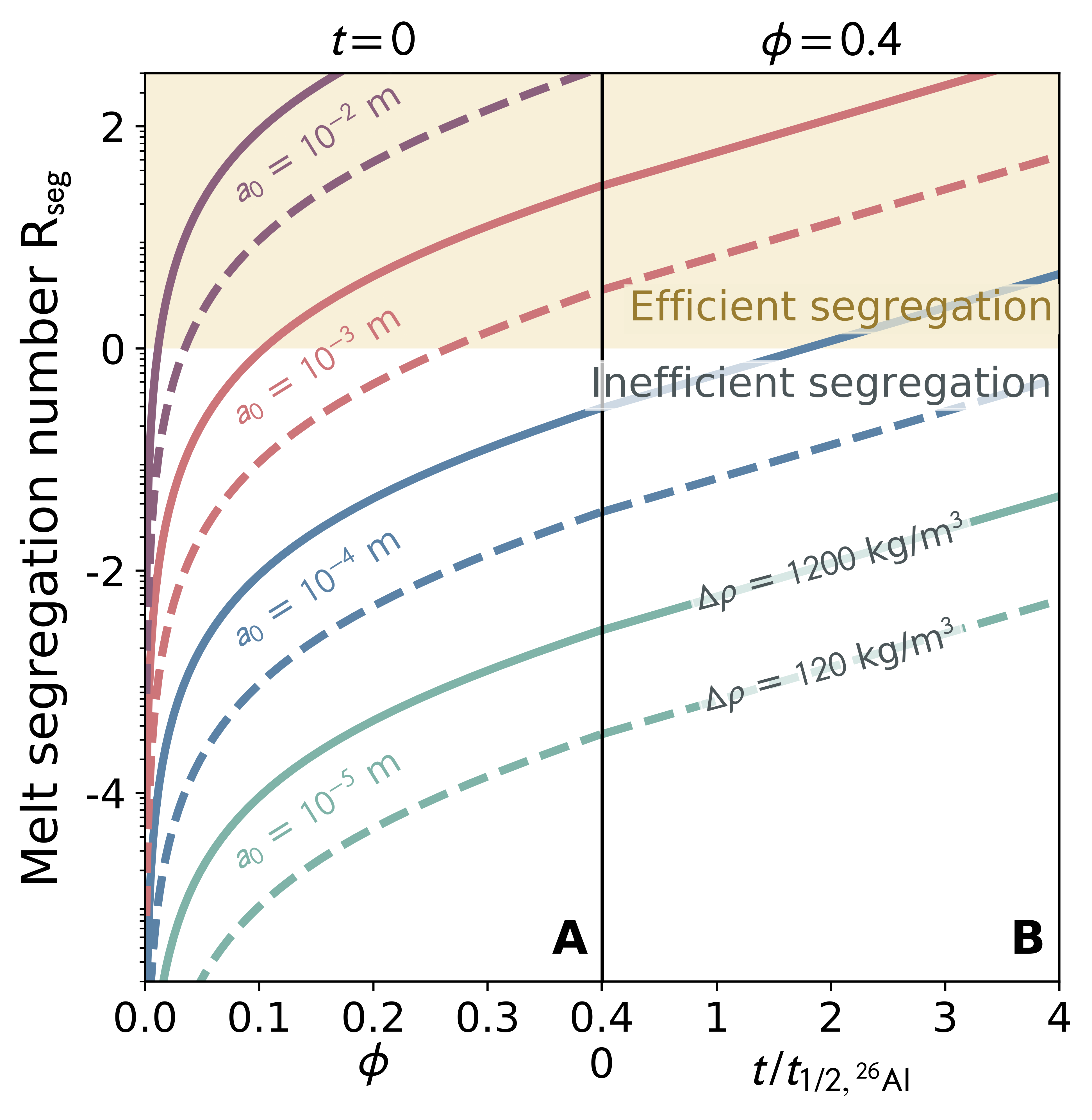}
    \caption{Scaling analysis of melt segregation propensity, with melt segregation number $\mathrm{R}_{\mathrm{seg}}$ = log$_{\mathrm{10}}$ ($\tau_{\mathrm{heat}}$/$\tau_{\mathrm{mt}}$). (A) At $t$ = 0 Myr (CAI formation) and with rising melt fraction $\phi$, the migration velocity increases, and so the system is more likely to become segregated. At around $\phi \gtrsim 0.4$--$0.6$, the magma ocean regime is reached and the system would be dominated by turbulent convection. (B) For fixed melt fraction $\phi = 0.4$ and later times ($=$ weaker radiogenic heating) the melt segregation number rises further.}
    \label{fig:scaling_analysis}
\end{figure}
which quantifies the propensity of a planetesimal to undergo substantial melt segregation during the internal heating by $^{26}$Al as a function of the key model parameters. To anticipate the expected melt segregation regimes in planetesimals, we calculate $\mathrm{R}_{\mathrm{seg}}$ for a reasonable range of melt fractions ($\phi_0 \in [0,0.4]$) below the rheological transition \citep[$\phi_{\mathrm{trans}}$ $\approx 0.4$--$0.6$,][]{2009GGG....10.3010C}, formation times ($t_0 \in [0,4]$ Myr), grain sizes ($a_0 \in [10^{-5},10^{-2}]$ m) and melt-rock density contrasts ($\Delta \rho_0 \in [120,1200]$ kg m$^{-3}$). Figure \ref{fig:scaling_analysis} shows that a growing melt fraction increases the melt segregation number through its effect on permeability. Larger grain sizes and higher density contrasts also significantly enhance segregation, but the effect of the latter is less pronounced than the former. In Figure \ref{fig:scaling_analysis}b, holding melt fraction fixed, the heating rate decreases with later formation times, which again serves to favor melt segregation relative to melt production.

From this scaling analysis, we conclude that melt segregation can in principle occur on a time scale that is relevant compared to the internal heating by $^{26}$Al in planetesimals \citep[e.g.,][]{2006M&PS...41...95H,2016Icar..274..350L,2018Icar..302...27L}. However, it crucially depends on the dynamic evolution of the melt fraction, which is controlled by the fluid mechanics of melt transport in a deforming rock matrix, and the thermo-chemical evolution of the body. In particular, Figure \ref{fig:scaling_analysis} highlights the importance of considering the evolution of internal heating and melt segregation over time. The scaling analysis does not yet capture any time-dependent effects of interest here, which include the potential accumulation of magma beneath a primitive lid and the redistribution of the heat source by transport of melt enriched in $^{26}$Al. In order to assess these dynamic processes we require a time-dependent evolution model, which we introduce in the next section. 

\section{Method}
\label{sec:methods}

\begin{table*}[hbt!]
\centering
\caption{Scaling quantities, definitions and parameter values introduced in the scaling analysis, R\_DMC and two-phase flow models. Varying model parameters are named in the text and figures. Parameters not listed here are as given in Table 1 in \citet{KellerKatz2016}.}
\vspace{0.2cm}
{\scriptsize
\begin{tabular}{lllll}
\hline
Parameter & Symbol & Unit & Value\\
\hline
Geometric factor & $b$ & non-dim. & 100\\
Melt fraction exponent & $n$ & non-dim. & 3\\
Solid fraction exponent & $m$ & non-dim. & 2\\
Melt shear viscosity & $\mu$ & Pa s & 1\\
Thermal expansivity & $\alpha$ & K$^{-1}$ & $3 \times 10^{-5}$\\
Specific heat capacity & $c_{\mathrm{p}}$ & J kg$^{-1}$ K$^{-1}$ & 1100\\
Thermal diffusivity & $\kappa$ & m$^{2}$ s$^{-1}$ & $1.14 \times 10^{-6}$\\
{\em olv} initial mass fraction & $\bar{c}^\mathrm{olv}$ & wt \% & 50 \\
{\em pxn} initial mass fraction & $\bar{c}^\mathrm{pxn}$ & wt \% & 35  \\
{\em fsp} initial mass fraction & $\bar{c}^\mathrm{fsp}$ & wt \% & 15  \\
{\em olv} melting point & $T_{\mathrm{m,0}}^{\mathrm{olv}}$ & K & 2050 \\
{\em pxn} melting point & $T_{\mathrm{m,0}}^{\mathrm{pxn}}$ & K & 1500 \\
{\em fsp} melting point & $T_{\mathrm{m,0}}^{\mathrm{fsp}}$ & K & 1350 \\
Entropy gain of fusion & $\Delta S$ & J K$^{-1}$ & 320  \\
Curvature coefficients & $r^{\mathrm{olv}}$, $r^{\mathrm{pxn}}$, $r^{\mathrm{fsp}}$ & J kg$^{-1}$ K$^{-1}$ & 50, 20, 10  \\
Linear $P$-coefficients & $B^{\mathrm{olv}}$, $B^{\mathrm{pxn}}$, $B^{\mathrm{fsp}}$ & K GPa$^{-1}$ & 60, 100, 120  \\
Rock density & $\rho_{\mathrm{0}}$ & kg m$^{-3}$ & 3200 \\
Reference rock viscosity & $\eta_{\mathrm{0}}$ & Pa s & $10^{19}$ \\
Shear viscosity cut-off & $\eta_{\mathrm{min}}$ & Pa s & $10^{16}$  \\
Compaction viscosity cut-off & $\zeta_{\mathrm{min}}$ & Pa s & $10^{17}$  \\
Initial temperature & $T_{\mathrm{init}}$ & K & 290  \\
Surface temperature & $T_{\mathrm{space}}$ & K & 290  \\
Planetesimal radius & $R_{\mathrm{P}}$ & km & 60 \\
Grain size & $a$ & m & [$10^{-5}$, $10^{-2}$] \\
Formation time & $t_{\mathrm{form}}$ & Myr & [0, 4] \\
Melt-rock density contrast & $\Delta \rho$ & kg m$^{-3}$ & [120, 1200] \\
\hline
\end{tabular}
 }

\label{tab:param}
\end{table*}

\subsection{Melting and heat source partitioning} 
\label{sec:rdmc}

Studies of primitive meteorites \citep{1991Icar...90..107M,2010M&PS...45..123D} and equilibrium condensation sequence calculations \citep{2000GeCoA..64..339E} suggest that the main rock-forming mineral phases in solar system planetesimals were olivine (dominantly forsterite, Mg$_{2}$SiO$_{4}$), pyroxene (dominantly enstatite, MgSiO$_{3}$), and feldspar (dominantly anorthite, CaAl$_{2}$Si$_{2}$O$_{8}$). Ignoring minor contributions from CAIs and trace minerals, feldspar represents the major host phase of $^{26}$Al in rocky planetesimals. In addition to the timing of accretion and size of a planetesimal \citep[e.g.,][]{2016Icar..274..350L,2018Icar..302...27L}, magma genesis depends on the relative abundance of these phases and the concentration of volatiles. However, to avoid further complexity, we will consider only dry melting here, which is justified if volatile degassing during accretion is efficient \citep[e.g.,][]{Monteux2018}. We therefore formulate a model for melting and melt-rock partitioning of these major mineral phases.

We employ the \textsc{r\_dmc} method of \citet{KellerKatz2016} to calculate an idealized thermodynamic equilibrium at given temperature, pressure, and bulk composition, and linear kinetic reaction rates for a multi-component system. We define three compositional pseudo-components and their mass-concentrations in the solid (rock), $c^i_s$, and liquid (melt), $c^i_\ell$. These capture the leading-order behavior of classes of minerals grouped by similar melting points and partitioning behavior: {\em olv} (olivine-like, $i=1$), {\em pxn} (pyroxene-like, $i=2$) and {\em fsp} (feldspar-like, $i=3$). The mass fraction of all three components must sum to unity in both phases.

Using a simplified form of ideal solid solution theory \citep{2011GeoJI.184..699R}, we determine the component partition coefficients at given $P,T$-conditions, 
\begin{linenomath*}
\begin{align}
	K^i = \frac{c_{s}^{i,\mathrm{eq}}}{c_\ell^{i,\mathrm{eq}}} = \mathrm{exp} \left[ \frac{L^i}{r^i} \left( \frac{1}{T} - \frac{1}{T_{\mathrm{m}}^i (P)} \right) \right] \: ,
\end{align}
\end{linenomath*}
which are the ratios of solid, $c_{s}^{i,\mathrm{eq}}$, to liquid, $c_\ell^{i,\mathrm{eq}}$, component concentrations at equilibrium. The latent heat of pure-component fusion,
\begin{linenomath*}
\begin{align}
L^i = \Delta S T_{\mathrm{m,0}}^i \: ,
\end{align}
\end{linenomath*}
is given by the entropy gain of fusion, $\Delta S$, and the pure-component melting temperatures at zero pressure, $T_{\mathrm{m,0}}^i$. Curvature coefficients $r^i$ adjust the temperature dependence of partition coefficients. The pressure-dependent pure-component melting points are parameterized as
\begin{linenomath*}
\begin{align}
T_{\mathrm{m}}^i (P) = T_{\mathrm{m,0}}^i + B^i P \: ,
\end{align}
\end{linenomath*}
with linear slopes $B^i$. 

At a given volume-averaged bulk composition (lever rule)
\begin{linenomath*}
\begin{align}
\bar{c}^i = (1-\phi) c_{s}^i + \phi c_\ell^i \: , 
\end{align}
\end{linenomath*}
we numerically determine the equilibrium melt fraction $\phi^{\mathrm{eq}}$ that satisfies the partitioning coefficients $K^i$ by combining the lever rules with the unity sum constraint on all components
\begin{linenomath*}
\begin{align}
\sum_{i=1}^{n} \frac{\bar{c}^i}{\phi^{\mathrm{eq}}/K^i + (1 - \phi^{\mathrm{eq}})} = \sum_{i=1}^{n} \frac{\bar{c}^i}{\phi^{\mathrm{eq}} + (1 - \phi^{\mathrm{eq}})K^i}. \label{eq:unitysum}
\end{align}
\end{linenomath*}
From $\phi^{\mathrm{eq}}$, we then calculate the equilibrium phase compositions for solid and melt,
\begin{linenomath*}
\begin{subequations}
\begin{align}
c_{s}^{i,\mathrm{eq}} & = \frac{\bar{c}^i}{\phi^{\mathrm{eq}}/K^i + (1 - \phi^{\mathrm{eq}})} \: ,\\
c_\ell^{i,\mathrm{eq}} & = \frac{\bar{c}^i}{\phi^{\mathrm{eq}} + (1 - \phi^{\mathrm{eq}})K^i} \: .
\end{align}
\end{subequations}
\end{linenomath*}

\begin{figure}[htb!]
    \centering
    \includegraphics[width=0.45\textwidth]{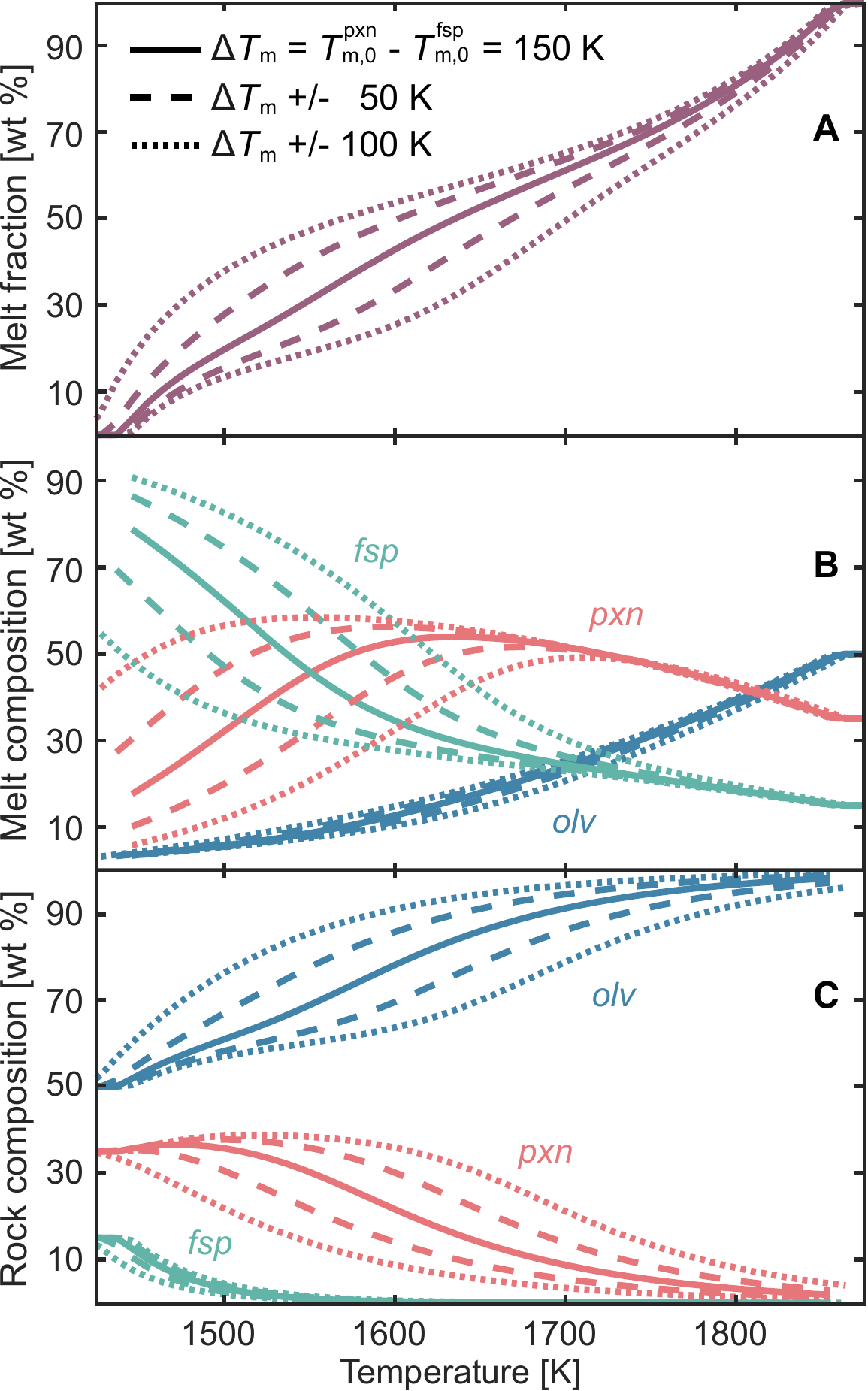}
    \caption{Temperature dependence of melt fraction (A), melt (B) and rock (C) composition, with varying melting point difference between {\em fsp} and {\em pxn}, $\Delta T_{\mathrm{m}}$ = $T_{\mathrm{m}}^{\mathrm{pxn}}$ - $T_{\mathrm{m,0}}^{\mathrm{fsp}}$, which changes the partitioning coefficient of {\em fsp} and the composition of the earliest melts. If the {\em fsp} melting point is close to the one of {\em pxn}, the initial melt composition is close to a 50--50 mixture. For higher melting point difference, the first melts are dominated by {\em fsp}, and thus the heating component ($^{26}$Al) of the system preferentially follows the dynamics of the earliest melts. When the absolute temperature of the system further rises and approaches the {\em olv} melting point, the composition converges towards the initial pure solid setting.} %
    \label{fig:rdmc1}
\end{figure}

Dynamic changes in pressure, temperature or bulk composition over time create disequilibrium. The mass transfer of component $i$ from solid to liquid that drives the system back towards equilibrium is assumed to occur at linear kinetic rates,
\begin{linenomath*}
\begin{align}
\Gamma^i = \mathcal{R} \left( \phi^{\mathrm{eq}} c_\ell^{i,\mathrm{eq}} - \phi c_\ell^i \right) \: ,
\end{align}
\end{linenomath*}
with a rate factor $\mathcal{R} = \rho_{\mathrm{0}} / \tau_{\Gamma}$, which restores equilibrium over a time scale $\tau_{\Gamma}$ at reference density $\rho_{\mathrm{0}}$. The sum of all component reaction rates is the total melting rate $\Gamma = \sum_i \Gamma^i$. All parameter values are given in Table \ref{tab:param}.

We first consider the aluminum partitioning behavior in a closed, isobaric ($P=0$) system under increasing temperature. For consistency with earlier work \citep{2013NatGe...6...93T,2014MPS...49.1083G,2016Icar..274..350L,Monteux2018,2018Icar..302...27L,Hunt18EPSL}, we choose the lowest component melting point, $T^{\mathrm{fsp}}_{\mathrm{m,0}} = 1350$ K, to conform with previous estimates of the silicate solidus. To test different partitioning behaviors, we vary the relative temperature difference between the melting points for the $^{26}$Al-hosting {\em fsp} and the {\em pxn} components, $\Delta T_{\mathrm{m}} = T^{\mathrm{pxn}}_{\mathrm{m,0}} - T^{\mathrm{fsp}}_{\mathrm{m,0}}$, with $T^{\mathrm{pxn}}_{\mathrm{m,0}} \in$ [1400, 1600] K. The resulting rock and melt compositions in the ternary system {\em olv}-{\em pxn}-{\em fsp} are shown in Figure \ref{fig:rdmc1}. The larger $\Delta T_{\mathrm{m}}$ between {\em fsp} and {\em pxn}, the more incipient melt will be enriched in {\em fsp}, and thus $^{26}$Al. As the degree of melting increases with temperature, more {\em pxn} is dissolved into the melt. Finally, in absence of melt migration, the melt composition would converge to the bulk composition as the system approaches complete melting. At our chosen reference calibration $T^{\mathrm{pxn}}_{\mathrm{m,0}} = 1500$ K, the melt initially comprises $\approx 80 \%$ {\em fsp}, but becomes relatively enriched in {\em pxn} by the time the melt fraction reaches $\phi_{\mathrm{trans}}$.

\subsection{Two-phase, multi-component fluid model}
\label{sec:two-phase}
As partial melts segregate from their residual, the interior becomes gradually depleted of {\em fsp} and thus its heat source $^{26}$Al. To model this dynamic process, we couple the multi-component melting model above to the two-phase reactive transport model of \citet{KellerKatz2016}. The fluid mechanics part of the model is based on \citet{1984JPet....5..713M}. Here, we give a brief summary of the governing equations and constitutive relations.

The physical model is derived from statements for the conservation of phase and component mass, phase momentum, and total energy. We consider the model in a Cartesian coordinate system with gravity pointing down the vertical coordinate, $\mathbf{g} = -g \mathbf{\hat{z}}$. The governing equations are formulated in the geodynamic limit (liquid viscosity $\mu \ll$ rock viscosity $\eta$), using the extended Boussinesq approximation (densities $\rho_{s} = \rho_\ell = \rho_0$ taken equal and constant except when multiplying gravity). The fluid mechanics governing equations are
\begin{linenomath*}
\begin{subequations}
\begin{align}
  \nabla \cdot \eta \left[\nabla\mathbf{v}_{s} + (\nabla \mathbf{v}_{s})^T - \frac{1}{3} \mathbf{I}\nabla\cdot\mathbf{v}_{s}\right] - \nabla P_\ell + \phi \Delta \rho \mathbf{g} = \nabla P_\mathrm{C} & \: , \label{eq:fluid1} \\
-\nabla \cdot \mathbf{v}_{s} = \nabla \cdot \mathbf{v}_\mathrm{D} & \: , \label{eq:fluid2} \\
\mathbf{v}_\mathrm{D} = \phi (\mathbf{v}_\ell - \mathbf{v}_s) = -\frac{k_{\phi}}{\mu} \left(\nabla P_\ell + \Delta \rho \mathbf{g} \right) & \: , \label{eq:fluid3} \\
P_\mathrm{C} = (1-\phi)(P_s - P_\ell) = - \zeta \nabla \cdot \mathbf{v}_s & \: . \label{eq:fluid4}
\end{align}
\end{subequations}
\end{linenomath*}
They are posed in four independent variables, the dynamic pressures, $P_s$, $P_\ell$, and velocities $\mathbf{v}_s$, $\mathbf{v}_\ell$, of the solid and liquid phases. Two dependent variables, the Darcy segregation flux, $\mathbf{v}_\mathrm{D}$, and the compaction pressure, $P_\mathrm{C}$, express the mechanical interactions between the phases. If these vanish, the equations become identical with the Stokes system.  Assuming a diffusion creep rheology with melt-weakening, the shear viscosity of the rock matrix is given by
\begin{linenomath*}
\begin{align}
\eta & = A_0 \, a_0^3 \, \mathrm{exp} \left( \frac{E_{\mathrm{a}} + P V_{\mathrm{a}}}{RT} - \lambda \phi  \right),
\end{align}
\end{linenomath*}
with prefactor $A_0$, activation energy, $E_{\mathrm{a}}$, and activation volume $V_{\mathrm{a}}$ as in \citet{hirth2003rheology} and \citet{2002E&PSL.201..491M}. $R$ is the universal gas constant, and $\lambda \approx 30$ the melt-weakening factor. Permeability is set by the Kozeny-Carman relation (Eq.~\ref{eq:permeability}), with powerlaw exponents $n=3$, $m=2$ for the melt and solid fractions, respectively. The compaction viscosity is set to $\zeta = r_{\zeta} \eta / \phi$, with $r_\zeta \geq 1$ the shear to compaction viscosity ratio.

To these equations we add thermo-chemical evolution equations for temperature, $T$ (assuming thermal equilibrium between phases), melt fraction, $\phi$, and pseudo-component concentrations in the solid and liquid phases, $c^i_s$ and $c^i_\ell$:
\begin{linenomath*}
\begin{subequations}
\begin{align}
\frac{\bar{\mathrm{D}}T}{\mathrm{D}t} &= \kappa \nabla^2 T - \sum_{i=1}^{n} \dfrac{L^i\Gamma^i}{\rho_0 c_\mathrm{p}} + \dfrac{\Psi}{\rho_0 c_\mathrm{p}} + \dfrac{H_{^{26}\mathrm{Al}}(t)}{c_\mathrm{p}} + \dfrac{\alpha T}{c_{\mathrm{p}}} g w_z \: , \label{eq:energy} \\
\frac{\mathrm{D}_{s} \phi}{\mathrm{D} t} &= (1-\phi) \nabla \cdot \mathbf{v}_s + \Gamma/\rho_{0} \: , \label{eq:melt} \\
\frac{\mathrm{D}_\ell c_\ell^i }{\mathrm{D} t} &= \hspace{11pt} \dfrac{\Gamma^i - c_\ell^i \Gamma}{\phi \rho_0} \: , \label{eq:comp_s} \\
\frac{ \mathrm{D}_{s} c_{s}^i}{\mathrm{D} t} &= - \dfrac{\Gamma^i - c_{s}^i \Gamma}{(1-\phi)\rho_0 } \: . \label{eq:comp_l}
\end{align}
\end{subequations}
\end{linenomath*}

Temperature evolves due to advection, thermal diffusion, latent heat exchange of reactions, heating by viscous dissipation $\Psi$ and internal heating $H_\mathrm{^{26}Al}(t)$, and adiabatic decompression. Melt fraction evolves due to advection, compaction and reactions, and composition by advection and reaction. The material derivative of the two-phase mixture is $\bar{\mathrm{D}}/\mathrm{D}t = (1-\phi) \mathrm{D}_s/\mathrm{D}t + \phi \mathrm{D}_\ell/\mathrm{D}t$, with $\mathrm{D}_s/\mathrm{D}t = \partial/\partial t + \mathbf{v}_{s} \cdot \nabla$, and $ \mathrm{D}_\ell/\mathrm{D}t = \partial/\partial t + \mathbf{v}_\ell \cdot \nabla$. $\kappa$ is the thermal diffusivity, $w_{z} = [(1 - \phi) \mathbf{v}_{s} + \phi \mathbf{v}_\ell] \cdot \hat{\mathbf{z}}$ the vertical bulk speed, and $\alpha$ the thermal expansivity.

\begin{figure*}[tbh!]
    \centering
    \includegraphics[width=0.90\textwidth]{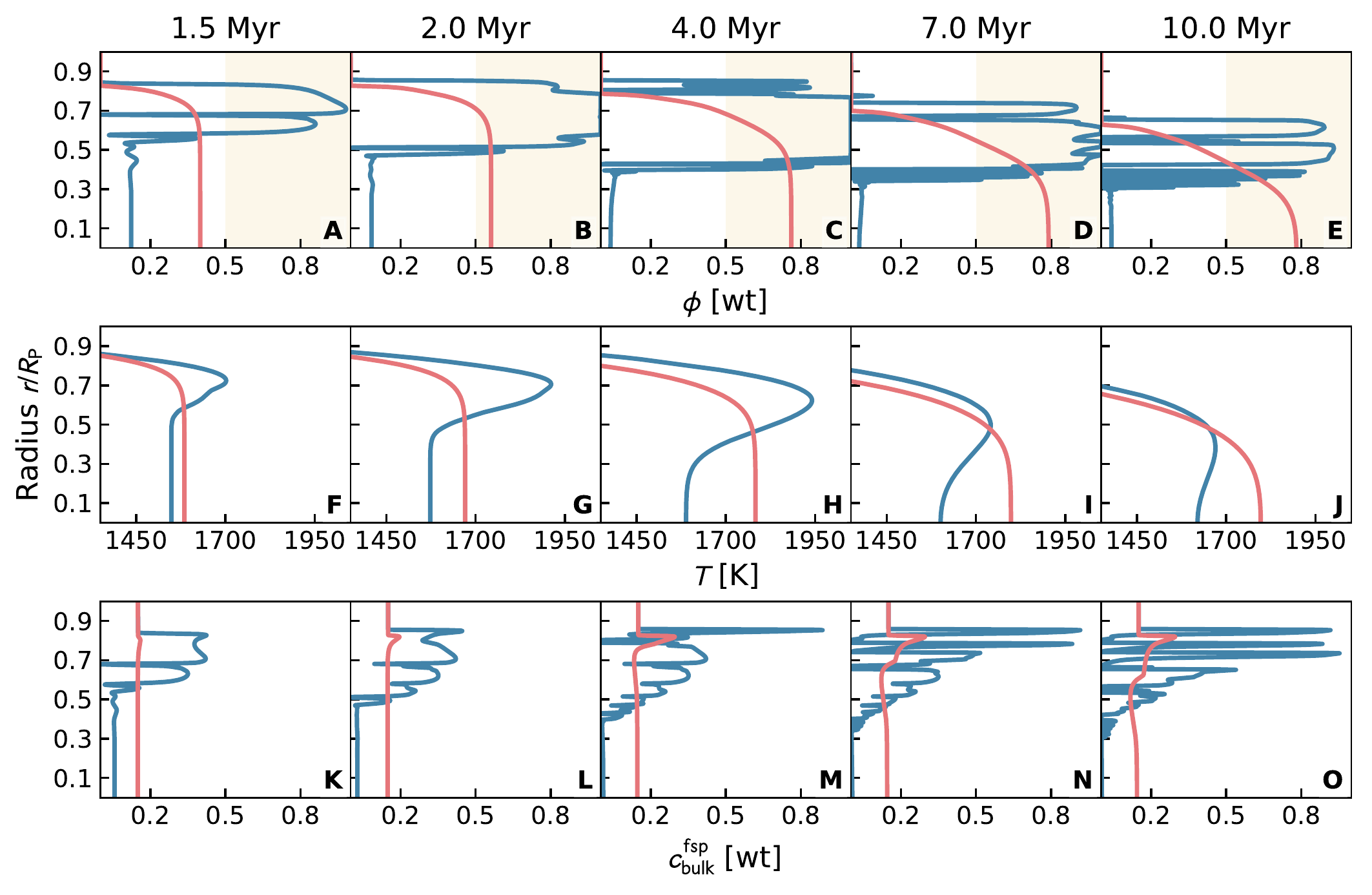}
    \caption{Time evolution of melt fraction (A--E), temperature (F--J) and \emph{fsp} bulk composition (K--O) for two end-member models with radii $R_{\mathrm{P}}$ = 60 km and formation time $t_{\mathrm{form}}$ = 1.25 $\times$ $t_{1/2,^{26}\mathrm{Al}}$ = 0.90 Myr after CAIs. Red lines show a \emph{magma ocean} model with $\Delta \rho$ = 200 kg m$^{-3}$ and $a_0$ = $10^{-4}$ m, blue lines a 
    \emph{magma sill} model with $\Delta \rho$ = 800 kg m$^{-3}$ and $a_0$ = $10^{-3}$ m. Upon progressive heating, the \emph{magma sill} model builds up melt accumulates below the cold upper lid, depleting the center of the planetesimal of silicate melts. High melt fractions $\phi$ $>$ $\phi_{\mathrm{crit}}$ $:=$ 0.5 (yellow areas) are only reached in the sub-lid sills. {\em fsp} enrichment in the sill structure leads to a temperature inversion of $\approx$ 400 K at peak melting. The \emph{magma ocean} model, in contrast, shows a near-isothermal internal temperature and thus constant melt fraction structure in the interior. The {\em fsp} component shows notable deviations from the initial bulk composition only after $t \geq$ 2 Myr, when most of the $^{26}$Al is already decayed. A video showing the time evolution of the major thermo-chemical parameters and composition is linked to in the \nameref{sec:suppl}.}
    \label{fig:grid_sill}
\end{figure*}

The governing equations for the fluid mechanics (Eqs.~\ref{eq:fluid1}--\ref{eq:fluid4}) and thermo-chemistry (Eqs.~\ref{eq:energy}--\ref{eq:comp_l}) are discretized using the finite difference method on a rectangular, staggered grid and solved by two coupled Newton-Krylov solvers. The simulation software uses parallel data structures and solvers provided by \textsc{PETSc} \citep{petsc-efficient,katz2007PETSC}. Nonlinearities between the fluid mechanics and thermo-chemical sub-problems are resolved using a Picard fixed-point iterative scheme. During every iteration, Equation \ref{eq:unitysum} is solved using Newton's method to update the equilibrium melt fraction. The adopted model is strictly valid only for melt transport by porous flow below the disaggregation threshold. However, we cannot avoid models producing regions with higher melt content. To ensure that the equations do not produce numerically unstable solutions in these regions, we apply lower viscosity cut-offs to the shear viscosity ($\eta_\mathrm{num} = \eta + \eta_\mathrm{min}$) and compaction viscosity ($\zeta_\mathrm{num} = \zeta + \zeta_\mathrm{min} / (1-\phi)$). The effect of this regularization is to dampen the segregation velocity and compaction pressure at elevated $\phi$. As a result, our model will produce stable solutions above the rheological transition, but will underestimate chemical mixing and heat transport by rapid convection in the crystal-bearing suspensions that characterize this limit.

\subsection{Model setup and parameter space} 
\label{sec:setup}
We model magma genesis and transport along a 1D Cartesian column from the center to the surface of an initially homogeneous and isothermal body of 60 km radius. Planetesimals of this size are qualitatively representative of the interior evolution of planetesimals of $R_{\mathrm{P}}$ $\gtrsim 50$ km as these are dominated by a relatively large adiabatic interior and a thin ($\lesssim 10$ km) conductive lid, whereas the relative dimensions of these domains vary significantly for planetesimal radii $\lesssim 50$ km \citep[cf.][]{castillorogez2017,2018Icar..302...27L,Hunt18EPSL}. The computational domain includes 500 grid cells for a spatial resolution of 120 m. The surface boundary is $T_{\mathrm{space}} = 290$ K, similar to the temperature in the protoplanetary disk inside of the water snowline, while the center boundary is insulating, $\partial T / \partial z \mid_{\mathrm{z=0}} = 0$. We assume the accreted body is initially at ambient temperature, $T_{\mathrm{init}} = T_{\mathrm{space}}$. As noted above, gravity decreases linearly from the surface gravity down to zero at the center.

Magma and rock composition are modeled in the three-component compositional space of {\em olv}, {\em pxn} and {\em fsp} (Section \ref{sec:rdmc}). We use component melting points as in Figure \ref{fig:rdmc1} (solid lines). The solid-melt density contrast is varied as $\Delta \rho \in$ [200, 1200] kg m$^{-3}$ between runs to reflect FeO content and thus density of the melt reflecting the planetesimal's oxygen fugacity \citep{2014E&PSL.390..128F,wilson_keil_2017}. Grain size $a_{0}$, which controls both the permeability and rock viscosity, is held constant during  calculations; we consider values $a_{0}$ $\in$ [10 $\mu$m, 1 cm], from chondrite matrix-like dust to pallasite-like crystal sizes (cf. Figure \ref{fig:scaling_analysis}). Heating is induced solely by $^{26}$Al in the {\em fsp} component, whose redistribution hence affects the local heating rate. The initial heating rate is varied from $H_{^{26}\mathrm{Al}}(0) \in [1.52, 0.19] \times 10^{-7}$ W kg$^{-1}$, reflecting planetesimal formation times $t_{\mathrm{form}}$ $\in [0, 3]$ $\times$ $t_{1/2,^{26}\mathrm{Al}}$ $=$ $ [0, 2.2]$ Myr after CAIs. To limit model complexity, we ignore the potential heat contribution from $^{60}$Fe \citep[see, e.g.,][]{2015ApJ...802...22T,2016MNRAS.462.3979L}.

\section{Results}
\label{sec:results}

\begin{figure*}[tbh!]
    \centering
    \includegraphics[width=0.80\textwidth]{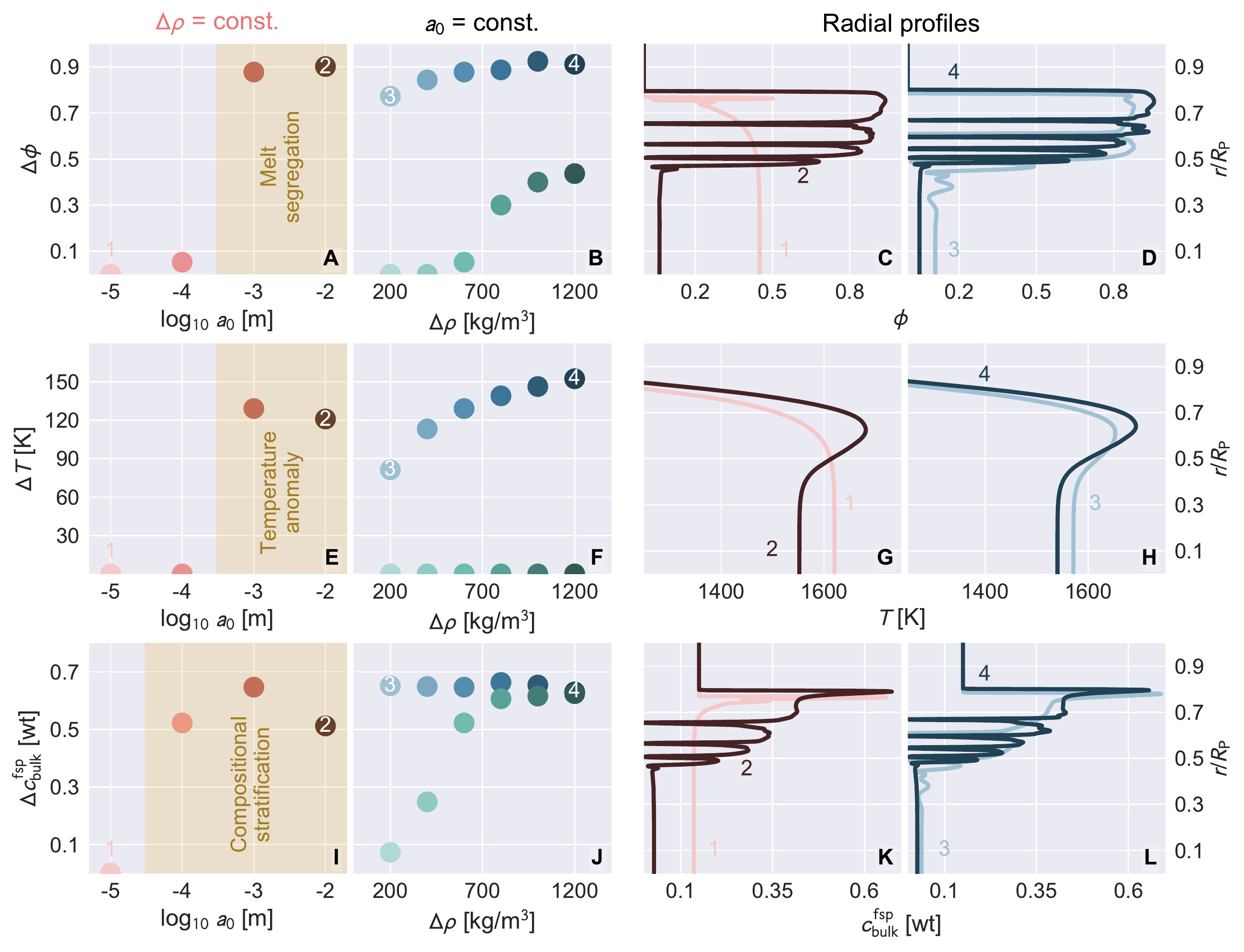}
    \caption{Parameter study of the influence of grain size $a_{0}$ and density contrast $\Delta \rho$ on melt segregation $\Delta \phi$, temperature inversion $\Delta T$, and compositional stratification $\Delta c_{\mathrm{fsp}}$, for planetesimals with $R_{\mathrm{P}}$ = 60 km, $t_{\mathrm{form}}$ = 1.5 $\times$ $t_{1/2,^{26}\mathrm{Al}}$. Panels (A,E,I) show the metric deviation for constant density contrast $\Delta \rho$ = 600 kg m$^{-3}$ and varying grain size $a_{0}$, indicating a steep gradient between grain sizes of $10^{-4}$ m and $10^{-3}$ m. For these two values fixed (blue: $a_{0}$ = $10^{-3}$ m, green: $a_{0}$ = $10^{-4}$ m), panels (B,F,J) display the metric deviations for varying $\Delta \rho$. Here, variations in density contrast are outweighed by those in grain size. Models with $a_{0}$ = $10^{-3}$ m feature notable melt segregation, temperature inversions, and compositional differentiation. Models with $a_{0}$ = $10^{-4}$ m only do so towards the high end of density contrasts, $\Delta \rho$ $\gtrsim$ 700 kg m$^{-3}$. Panels (C/D, G/H, K/L) show the radial profiles for the end-member models of the variations from (A/B, E/F, I/J). In general, variations in grain size $a_{0}$ outweigh effects from increasing density contrast $\Delta \rho$. \emph{Magma sill} structures only form for grain sizes $a_{0}$ $> 10^{-4}$ m.}
    \label{fig:params}
\end{figure*}

\begin{figure}[bth!]
    \centering
    \includegraphics[width=0.45\textwidth]{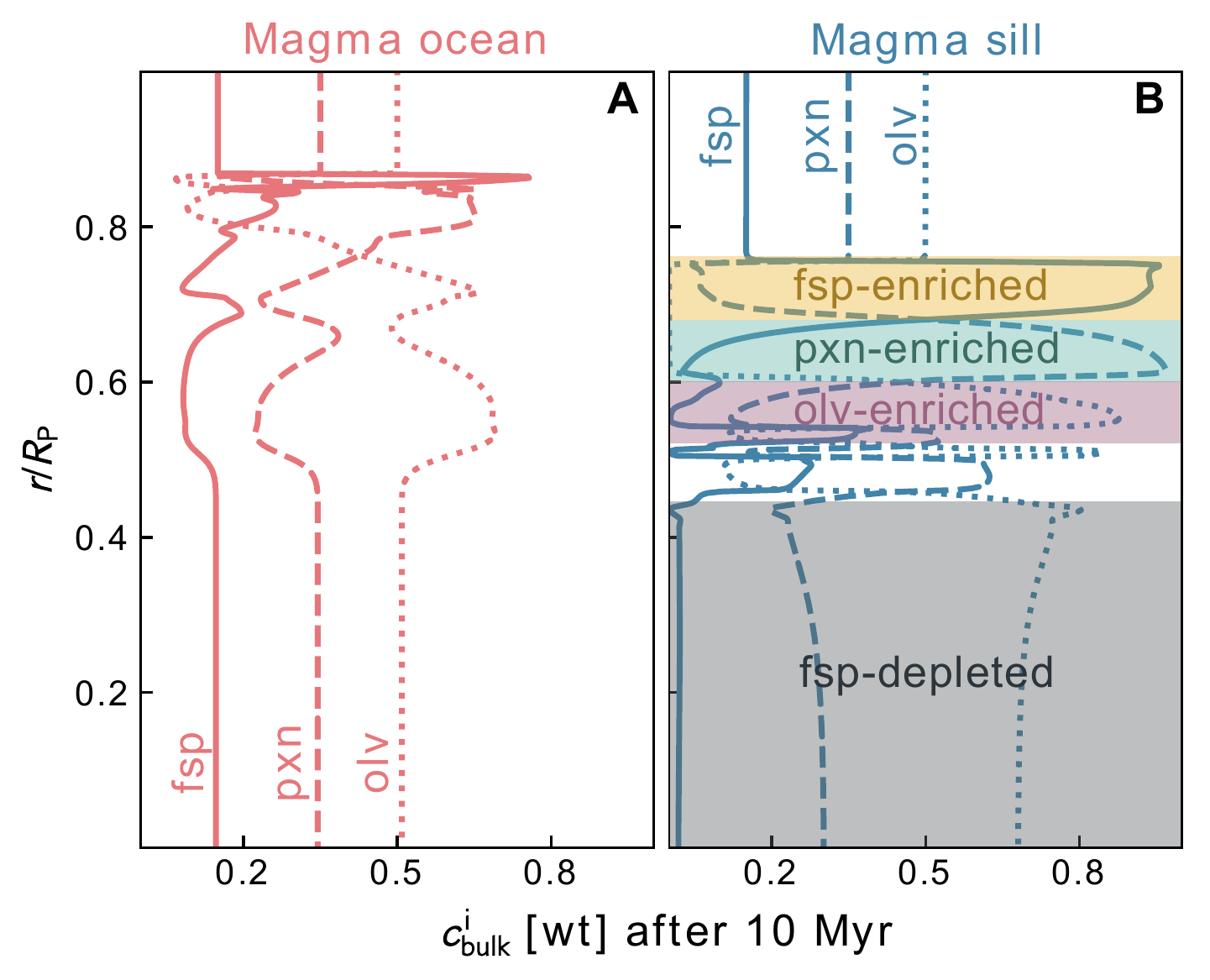}
    \caption{Compositional stratification after cooling and crystallization of magma beneath the primordial lid for \emph{magma ocean} (A) and \emph{magma sill}-type (B) models. \emph{Magma sill} cases with intermediate temperatures and thus high concentrations of {\em fsp} in the upper layers produce this signature.}
    \label{fig:cbulk}
\end{figure}

\begin{figure*}[htb!]
    \centering
    \includegraphics[width=0.95\textwidth]{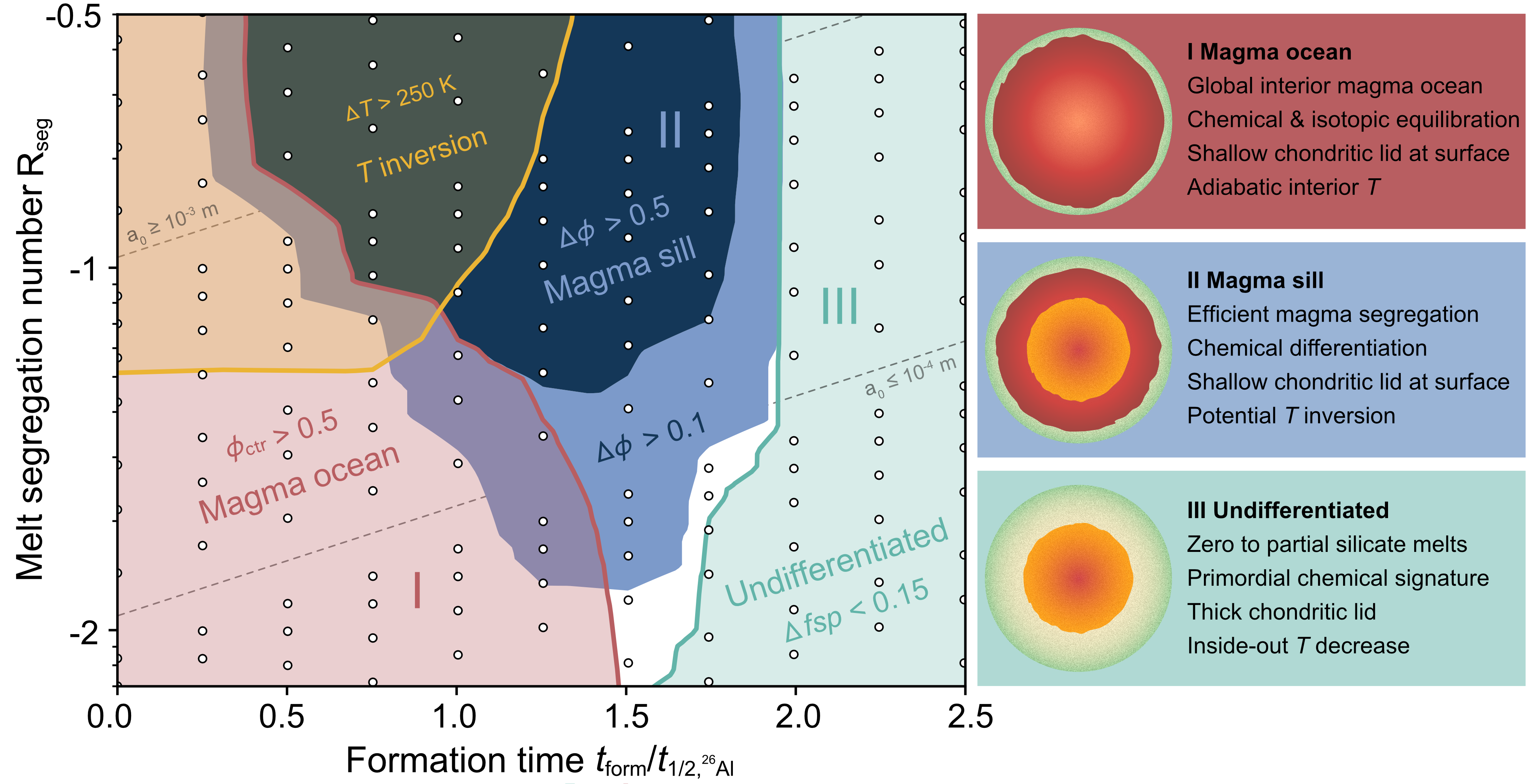}
    \caption{Evolution of silicate melt segregation with formation time $t_{\mathrm{form}}$ versus melt segregation number $\mathrm{R}_{\mathrm{seg}}$ at a reference melt fraction of $\phi_0 = 0.1$. Colormap values are plotted for the time of peak melting for each model (circles). We identify three primary melt dynamics regimes. (I) \emph{Magma ocean} models, where melting occurs more rapidly than melt migration, feature high melt fractions above the rheological transition in their center, $\phi_{\mathrm{ctr}}$ $>$ $\phi_{\mathrm{crit}}$ $:=$ 0.5. \emph{Magma ocean}-type evolution is preferred for early $t_{\mathrm{form}}$ and low $\mathrm{R}_{\mathrm{seg}}$, i.e., small $a_{0}$ and $\Delta \rho$. (II) \emph{Magma sill} models feature efficient melt segregation, with additional compositional stratification towards cooling-down of the planetesimals (cf. Figure \ref{fig:cbulk}). \emph{Magma sill}-type evolution is preferred for intermediate $t_{\mathrm{form}}$ $\approx$ 0.5--1.75 $\times$ $t_{1/2,^{26}\mathrm{Al}}$, and high $\mathrm{R}_{\mathrm{seg}}$. (III) \emph{Undifferentiated} models never show melt fractions $\phi_{\mathrm{ctr}}$ $>$ $\phi_{\mathrm{crit}}$, and never experience substantial compositional redistribution. They are preferred for late formation times $t_{\mathrm{form}}$ $\gtrsim$ 2.0 $\times$ $t_{1/2,^{26}\mathrm{Al}}$. In addition to these three regimes, we show the region of increasing temperature inversion in yellow.}
    \label{fig:magma_regime}
\end{figure*}

\subsection{Parameter study}
\label{sec:results1}

We find that model results across the tested parameter range fall into three qualitative categories. Some models show no substantial melting or segregation; we will not further discuss these \emph{undifferentiated} models here. The time evolution of the latter two categories shown in Figure \ref{fig:grid_sill} is generally the same: rapid initial heating leads to substantial melt production, followed by some degree of segregation, before ending with slow cooling from the top down. One category of models, which we identify as the \emph{magma ocean} end-member (red), evolve to where most of the interior is above the disaggregation threshold, whereas the other, the \emph{magma sill} end-member (blue), result in a melt-depleted interior beneath melt-rich sills (Fig. \ref{fig:grid_sill}A--E). The latter clearly shows a thermal inversion (Fig. \ref{fig:grid_sill}F--J) related to {\em fsp}-enrichment in the magma sills beneath the lid (Fig. \ref{fig:grid_sill}K--O).

The scaling analysis above predicts that grain size, $a_{0}$, and density contrast, $\Delta \rho$, are pertinent controls on melt segregation. Figure \ref{fig:params} shows the results of a detailed study of that parameter space. To quantitatively analyze the results, we introduce three metrics measuring the degree of
\begin{itemize}
\item melt segregation: $\Delta \phi$ = $\phi_{\mathrm{max}}$ -- $\phi_{\mathrm{ctr}}$, 
\item temperature inversion: $\Delta T$ = $T_{\mathrm{max}}$ -- $T_{\mathrm{ctr}}$,
\item {\em fsp} differentiation: $\Delta c_{\mathrm{fsp}}$ = $\overline{c}_{\mathrm{max}}^{\mathrm{fsp}}$ -- $\overline{c}_{\mathrm{ctr}}^{\mathrm{fsp}}$.
\end{itemize}
Here, $(\cdot)_\mathrm{max}$ denotes the maximum value in the computational domain and $(\cdot)_\mathrm{ctr}$ denotes the value at the base of the domain, i.e., the planetesimal center. With these metrics, we quantify the deviation from an interior structure with near-constant melt fraction, temperature, and bulk composition, as it would be expected in the absence of significant segregation.

Figure \ref{fig:params} shows the three metrics across a range of $a_{0}$ $\in$ [10$^{-5}$, 10$^{-2}$] m and $\Delta \rho$ $\in$ [200, 1200] kg m$^{-3}$. We find that grain size strongly controls the interior evolution of the planetesimals. If grain size remains below $a_{0}$ $< 10^{-4}$ m, melt density contrasts of $<500$ kg m$^{-3}$ are not sufficient to drive significant segregation. For density contrasts $>1000$ kg m$^{-3}$ and for larger grain sizes, melt segregation becomes significant, as evidenced by a step-increase in each of the three metrics. However, we find that since initially {\em fsp}-enriched melts migrate on time scales comparable or longer than $t_{1/2,^{26}\mathrm{Al}}$ ($\approx$ 0.72 Myr), the temperature inversion effect remains minor throughout (Figure \ref{fig:params}, panels B \& F).

\subsection{Silicate differentiation}
\label{sec:results2}

As the planetesimals cool and crystallize, \emph{magma sill} end-member cases exhibit a silicate differentiation trend towards compositional layering. In Figure \ref{fig:cbulk}, we compare a representative \emph{magma ocean} with a \emph{magma sill} model outcome. \emph{Magma ocean} models evolve towards a uniformly molten interior, with melt fractions above the rheological transition $\phi >$ $\phi_{\mathrm{trans}}$ and cool relatively unsegregated, such that the bulk concentrations of the {\em fsp}, {\em pxn} and {\em olv} components remain similar to the initial composition (Figure \ref{fig:cbulk}A). However, in \emph{magma sill} models, the melt-rich layers above a low-melt-fraction interior crystallize into distinct layers enriched in {\em fsp}, {\em pxn} and {\em olv}. This stratification reflects the component melting points ($T_{\mathrm{m}}^{\mathrm{fsp}}$ $<$ $T_{\mathrm{m}}^{\mathrm{pxn}}$ $<$ $T_{\mathrm{m}}^{\mathrm{olv}}$, Figure \ref{fig:cbulk}B). The melt-depleted central parts of the planetesimal are strongly depleted in {\em fsp} and somewhat less in {\em pxn}. In general, such compositional layering forms during the cooling stage and therefore does not cause substantial temperature inversion. The densities of the minerals represented by the pseudo-components suggest that such layering would be dynamically stable.

\subsection{Magma dynamics regimes}
\label{sec:results3}

Figure \ref{fig:magma_regime} shows the regimes of melt segregation and compositional stratification for different formation times, $t_{\mathrm{form}}$, and melt segregation numbers, $\mathrm{R}_{\mathrm{seg}}$. We quantify the boundaries separating the three characteristic regimes as follows. 
\begin{itemize}
\item[(I)] \emph{Magma ocean} regime: $\phi_{\mathrm{ctr}}$ $>$ $\phi_{\mathrm{crit}}$ $:=$ 0.5, where the whole interior melts to above the disaggregation threshold.
\item[(II)] \emph{Magma sill} regime: $\Delta \phi$ $>$ $\phi_{\mathrm{crit}}$, where melt segregation generates a melt-rich layer beneath the lid and depletes the interior of melt.
\item[(III)] \emph{Undifferentiated} regime: $\Delta c_{\mathrm{fsp}}$ $<$ $c_{\mathrm{bulk,0}}^{\mathrm{fsp}}$ = 0.15, where melting and melt segregation do not redistribute a substantial amount of {\em fsp}. 
\end{itemize}
In addition to these segregation and differentiation criteria, we show which models are most affected by substantial {\it temperature inversions}, $\Delta T$ $>$ 250 K. These inversions occur both for \emph{magma sill} and \emph{magma ocean} models and reflect rapid magma ascent on time scales shorter than $t_{1/2,^{26}\mathrm{Al}}$. We find that the \emph{magma sill} regime generally occurs at higher segregation numbers---at larger grain sizes or elevated density contrasts---and formation times less than $\approx$ 1.5 Myr after CAIs, with a peak at around 1 Myr after CAIs. Very early formation time, $t_{\mathrm{form}}$ $\leq$ 1 Myr, or lower melt segregation number, $\mathrm{R}_{\mathrm{seg}}$ $\lesssim$ -1.5, favor \emph{magma ocean}-type evolution. Formation later than $\approx$ 1.5 Myrs after CAIs results in limited melting and \emph{undifferentiated} planetesimals. {\it Temperature inversions} occur for $t_{\mathrm{form}}$ $\lesssim$ 1 Myr and $\mathrm{R}_{\mathrm{seg}}$ $\gtrsim$ -1.5. 

\section{Discussion}
\label{sec:discussion}

\subsection{Parametric controls on magma segregation} 

The models above present thermo-chemically coupled two-phase flow calculations that resolve the partitioning of the major rock-forming components and their transport by magma. Using this method, we show that magma segregation in planetesimals formed within 2 Myr after CAIs was potentially significant if magma ascent was rapid with respect to the rate of melt production controlled by $^{26}$Al decay. Expressed in terms of the melt segregation number $\mathrm{R}_{\mathrm{seg}}$, \emph{magma sill} models ($\Delta \phi > 0.5$) were produced for $\mathrm{R}_{\mathrm{seg}}$ $\geq$ -1.5. This requires that (i) the average grain size in these planetesimals was comparable to or larger than chondrules, on the order of $a_{0}$ $\geq 10^{-3}$ m, or (ii) in reducing environments \citep[iron-wüstite IW-2.5, or $\Delta \rho$ $\approx$ 1200 kg m$^{-3}$, respectively,][]{1984GeCoA..48..111B,2014E&PSL.390..128F} at $a_{0}$ $\geq 10^{-4}$ m. Moreover, we find that even in the case of significant magma redistribution into shallow sills, the segregation time scale is comparable to the evolutionary time scale of the protoplanetary disk and thus the accretion time scale. Therefore, scaling analysis alone (see Section \ref{sec:scaling}), does not adequately capture the time-dependent competition between melting, partitioning, and melt transport. Because of the protracted onset of melt ascent during heat-up, $^{26}$Al-hosting phases release at least parts of the $^{26}$Al decay energy in the deeper region of planetesimals, and substantial temperature inversions on the order of a few hundred $K$ are only observed for extreme cases with $\mathrm{R}_{\mathrm{seg}}$ $\gtrsim$ -1, or early formation times $t_{\mathrm{form}}$ $\lesssim$ 1 Myr with $\mathrm{R}_{\mathrm{seg}}$ $\gtrsim$ -2.

At $t_{\mathrm{form}}$ $= \tau_{^{26}\mathrm{Al}} \approx$ 1 Myr after CAIs the models show a clustering of \emph{magma sill} cases. This peak is due to the competition between heating and segregation, as introduced in Section \ref{sec:scaling}. For formation times earlier than $\approx$ 1 Myr, heating, and thus melting, is dominant, and new melt is generated faster than transported away. Around $t_{\mathrm{form}}$ $\approx$ 1 Myr, melt upwelling velocities become fast enough to exceed melt generation. Finally, at later times ($t_{\mathrm{form}}$ $\gtrsim$ 1.5 Myr after CAIs or 2 $\times$ $t_{1/2,^{26}\mathrm{Al}}$, respectively) melt production remains low and minimal to no segregation occurs.

Melt--rock density contrast is thought to be controlled by the oxidation state of the body. The chemical composition of accreting planetesimals is inherited from the oxidation state in the midplane of the protoplanetary disk. Hence, the time and place of formation is expected to influence the subsequent interior evolution as it relates to the effects of melt segregation. For example, planetesimals accreted towards the inner disk may feature lower oxygen fugacities and therefore higher $\Delta \rho$, compared to planetesimals accreted further out \citep{2015Icar..248...89R}. In our models, we find that the effect of density contrast on magma segregation is of secondary importance. If grain sizes were too small to allow for a sufficient permeability, variations in density contrast would not have led to significantly different outcomes. This finding is in contrast to previous studies \citep{2011MaPS...46..903M,2012ChEG...72..289W,2012Icar..217..339S,2014E&PSL.390..128F,wilson_keil_2017,fu2017}, which did not consider magnitude variations in grain size, or relied on grain growth by pure Ostwald ripening without taking into account mechanisms of growth inhibition and grain destruction \citep{2013Icar..224..126N,2014E&PSL.395..267N}.

\subsection{Implications for the role of grain size}

A main conclusion from our models is that the mean grain size $a_{0}$ is the dominant parameter controlling the magma transport rate inside planetesimals. There are two main mechanisms that affect grain size during planetesimal accretion and early evolution. First, grain sizes may differ depending on the orbital location and mineralogical composition \citep{2004Natur.432..479V} of the dust that agglomerates\footnote{Note that in astrophysics literature \textit{grain size} usually denotes the characteristic size of porous dust aggregates embedded in the disk flow, whereas we here use the term \textit{grain size}, $a_{0}$, for the characteristic size of mineral grains in a granular, polymineralic rock aggregate.} to form the planetesimals. Coagulated dust aggregates on the order of $a_{0}$ $\geq 10^{-4}$ m are in the critical size regime for fast radial drift towards the protostar, depending on orbital location and evolutionary state of the protoplanetary disk \citep{1977MNRAS.180...57W}. However, these sizes may facilitate planetesimal formation from local dust-gas interactions \citep{2015SciA....115109J,2016SSRv..205...41B} and can trigger planetesimal formation at various orbital separations and times \citep{2018A&A...614A..62D} with varying dust particle distributions within newly-formed bodies \citep{2017MNRAS.469S.149W}. Second, grain sizes may evolve during the heating and melting in the planetesimal interior after accretion. In this process, the grain size evolves due to competing mechanisms for growth and destruction \citep{2011GeoJI.184..719R,2016PEPI..253...31B}. Grain growth by either normal grain growth before the first melts arise, or Ostwald ripening during partial melting in a solid--liquid aggregate, is driven by the reduction of interfacial energies, and competes with size reduction due to the presence of secondary solid phases \citep{2010E&PSL.291...10H} and mechanical work due to solid-state deformation in planetesimal interiors \citep{2013NatGe...6...93T}. During later stages, when the melt fraction reaches or exceeds the disaggregation threshold, grain sizes are governed by the convective flow regime in the magma, leading to a variety of possible scenarios \citep{solomatov2015magma}.

As a comparison, grain sizes of meteorite classes span a wide range, from $\mu$m-sized dust to pallasite-like, cm-sized phenocrysts \citep{2004mete.book.....H}. Chondrites, generally regarded to be the most pristine rock samples from the early solar system, display a bimodal size distribution, split between $\mu$m-sized dust (`matrix') and chondrules, which show characteristic sizes of $\approx 10^{-4}$ m to $10^{-3}$ m \citep{2015ChEG...75..419F}, with drastically differing textures and mineral grain sizes. The ratio of chondrules to matrix varies greatly between different chondrite groups, resulting in complex mixtures and grain size distributions. Meteorites that likely underwent partial melting (`primitive achondrites'), like Acapulcoite-Lodranites, Winonites and Brachinites, display grain sizes around ~$10^{-4}$ m \citep{2004mete.book.....H,wilson_keil_2017}. Ureilites, Aubrites, and Angrites, which come from bodies with more extensive, or even body-wide, silicate melting feature larger grains, on the order of $10^{-3}$ m. However, these textures are the end result of million-year long evolutionary processes, and may have undergone repeated destruction and reaccretion cycles that reset their thermal histories and textures \citep{2018Icar..302...27L}. Therefore, based on the grain sizes observed in meteorites, we are strongly limited in assessing the grain size evolution in planetesimal interiors at the time of their first melting.

Interpretation of our results in this context suggests that the planetesimal interior evolution and the redistribution of chemical and isotopic heterogeneities during planetary accretion can be influenced by the planetesimal formation mechanism, its accretion location, and the local compositional distribution of grains in the protoplanetary disk. Further studies of planetesimal formation and mineralogical inventory are needed to link their formation processes to their subsequent dynamic evolution. The local grain-size distribution within planetesimals may be evolving during rapid gravitational collapse \citep{2017MNRAS.469S.149W} or more gradual growth \citep{2013A&A...557L...4K}. Also, ongoing accretion, for instance due to subsequent pebble accretion \citep{2016A&A...586A..66V}, may influence whether magma can reach the surface through fractures. Future experimental studies on grain size evolution of partially molten aggregates spanning the meteoritic compositional space will be needed to advance our understanding of melt migration in early solar system planetesimals.

\subsection{Implications for chemical differentiation}

Recently, it was shown that differentiation by percolation of Fe,Ni-FeS liquids in primordial planetesimals may not be complete and that at least some material remains trapped in the rock matrix \citep{2009PEPI..177..139B,2015E&PSL.417...67C}. But once silicate melting has reached the disaggregation threshold, the remaining metal droplets will rain out rapidly towards the forming core \citep{2018Icar..302...27L}. Therefore, even though the models in this study do not include a metal phase, they allow leading order predictions regarding core formation. 

In the \emph{magma ocean} case, we expect core formation to be rapid, with nearly complete loss of metals to the core. In the case of a \emph{magma sill}, a two-step process may occur. First, a small proto-core may form from incomplete percolation. Then, after the formation of the sill structure, the remaining metal within this region may rain out and accumulate at the interface between the melt-depleted deep interior and the \emph{magma sill} zone. This emerging metal pool will either percolate downwards or form diapirs sinking through the weakened partially molten rock. The thermo-mechanical evolution of such a two-step process needs to be tested by taking into account metal phases in self-consistent multi-phase flow models, in order to make robust predictions that can be compared to laboratory studies of the core formation process. \citet{Neumann2018} recently suggested a multi-stage core formation scenario for the IVB parent body, which is qualitatively consistent with the \emph{magma sill} regime we propose based on our models.

The limited melt segregation in our undifferentiated models may offer an explanation for the absence of remnant differentiated materials in the asteroid belt \citep{2012P&SS...66..137W,2013M&PS...48.2333M}. Conversely, chemical stratification resulting from melt segregation may help to explain the paucity of olivine-rich deposits on 4 Vesta's surface \citep{2014Natur.511..303C,2015Icar..254..190C,raymond2017}. Furthermore, the \emph{magma sill} models and the resulting chemical stratification we describe here are consistent with earlier work by \citet[][``shallow magma ocean'']{2013Icar..224..126N,2014E&PSL.395..267N} and \citet[][``completely liquid layer'']{mizzon2015magmatic}, predicting a subsurface layer of accumulated silicate melts below a cold lid resulting from efficient melt segregation \citep[cf. discussion in][]{raymond2017,Neumann2018}.

Finally, our results indicate that the importance of melt segregation in planetesimal interiors varied substantially and affected the redistribution of heat-producing elements, such as $^{26}$Al, during melt ascent. In the case of our \emph{magma sill} end-member models, we also expect other incompatible elements to be preferentially segregated to shallow layers of a planetesimal. The crustal stripping paradigm of planetary accretion assumes that frequent impacts during planetary growth eroded and redistributed significant amounts of shallow materials between planetesimals. The strongly variable degree of melt segregation, and the resulting range of variably differentiated major, trace, and isotopic compositions of shallow planetesimal layers could result in compositional differences between forming planets and chondritic meteorites \citep[e.g.,][]{2018E&PSL.484..276C}. Our simulations indicate that magma ascent governing material redistribution to the planetesimal crust occurs on a $\approx$ Myr time scale, i.e., comparable to the collisional time scale of planetary accretion. This suggests that models quantifying compositional deviations between planets and chondrites should take into account the evolution of planetesimal interiors during planetary accretion.

\subsection{Limitations \& future directions}

One of the limitations of our model is the use of a 1D Cartesian geometry, which introduces systematic errors as compared to a 1D spherical geometry assuming radial symmetry. Among these errors, our model over-predicts the heating-to-cooling ratio of planetesimals. Because we focus on planetesimals of 60 km radius that show a nearly isothermal evolution in the deep interior \citep[cf. Figure \ref{fig:grid_sill},][]{castillorogez2017,Hunt18EPSL}, the heat-up phase is consistent with a radial model. However, geometric errors result in an under-prediction of the rate of heat loss through the surface. Similar problems apply to melt segregation velocities and melting rates. For example, the liquid mass conservation equation, in spherical coordinates with radial symmetry, contains the term $2w_\ell/r$, which is neglected here (a similar term appears in the solid mass conservation equation).  For radially outward melt migration, this term has the effect of reducing ${\partial \phi}/{\partial t}$; its absence therefore leads to an over-prediction of melt fraction. To quantify the error introduced by neglecting this term we compared its size, computed \textit{a posteriori}, with the flux divergence ${\partial \phi w_{\mathrm{\ell}}}/{\partial z}$ that appears in the same equation. We made this comparison across models of the \emph{magma ocean} and \emph{magma sill} regimes. The results indicate that the geometrical error can reach a magnitude comparable to the divergence. However, throughout most of the domain and once \emph{magma sill} structures are established, the term is negligible. 

In this work we consider a diffusion creep rheology only. A more realistic rheology would be a composite of both diffusion and dislocation creep \citep[e.g.,][]{2016PEPI..253...31B,2018E&PSL.484..341M}. At the highest temperatures and grain sizes tested here, model behavior would likely fall into the dislocation creep regime, where the matrix viscosity becomes independent of grain size. In this case, the grain size sensitivity of the compaction length (Equation \ref{eq:compaction_length}) would decrease. 

Furthermore, we use a constant melt viscosity of 1 Pa\,s (Table \ref{tab:param}). However, silicate melt viscosity varies with temperature and composition \citep[e.g.,][]{2011MaPS...46..903M,mizzon2015magmatic}. For the compositional space explored here, a temperature dependent viscosity of 1--100 Pa\,s may be considered realistic. As we are interested in the consequences of melt segregation in planetesimals here, we chose a reasonable lower limit. In addition, because the ratio of permeability to melt viscosity controls the upwelling timescales (grain size squared versus linear melt viscosity, Equation \ref{eq:compaction_length}), the order of magnitude variability in grain size outweighs potential variations in melt viscosity.

As a consequence of our limiting assumptions on the geometry, rheology, and melt viscosity, the extent of the \emph{magma sill} regime in Figure \ref{fig:magma_regime} may be overestimated. More efficient cooling in a spherical geometry would lead to decreasing melt fractions and thus lower migration speeds (Figure \ref{fig:scaling_analysis}), as would a stronger heat flux from turbulent convection above the disaggregation threshold \citep{solomatov2015magma}, and the geometric influence in spherical or higher-dimensional geometry. A weaker dependence on grain size in the models with the highest temperatures and largest grain sizes would also decrease migration speeds. Therefore, while our models constrain the possible phase space of melt migration in early solar system planetesimals, more complex models not bound to the above limitations would result in a reduced stability field where melt segregation becomes dominant than shown in Figure \ref{fig:magma_regime}.

Our models do not feature a metal phase that would allow the direct resolution of metal percolation \citep{2017PNAS..11413406G,Neumann2018}, and therefore our results only allow for qualitative inferences about possible core formation scenarios. However, the complexities of textural equilibrium phase topologies are not yet fully understood \citep[cf.][]{2018RSPSA.47470639R}, in particular for systems with several immiscible liquid phases. For example, the wetting angles formed between metal liquids with silicate minerals in the presence of a silicate melt remains unclear, leaving open the debate around a possible percolation threshold for metal liquids at low melt fractions \citep{2015E&PSL.417...67C}. Further work needs to be undertaken to better quantify these effects.

Our choice of melting model in the form of a ternary ideal solution limits the degree to which the model may represent natural melting behavior. For example, our model does not reproduce the eutectic behavior expected for silicate compositions as the ones considered here \citep[see discussion in][]{KellerKatz2016}, nor does it include volatiles and incompatible elements producing low-degree, incompatible-enriched melts at temperatures below the anhydrous solidus. Using a more consistent petrological model that takes into account the non-ideal thermodynamics of the full range of major elements and mineral phases would likely lead to more complex relations between heating, melt production, and element partitioning \citep{2018arXiv180900079K}. Significant differences in aluminum partitioning, which is the focus here, are likely confined to the onset of melting, where low-degree, enriched melts may have important control on geochemical evolution. Moreover, our dry models neglect volatile-driven eruptions, which were previously discussed as a catalyst for upward transport via explosive volcanism \citep{fu2017,wilson_keil_2017}. If substantial volatile quantities could be retained, this mechanism would decrease the retention of magma in the upper layers of planetesimals and potential chemical stratification upon crystallization of the silicate material. However, the relatively low pressure at the planetesimal sizes we consider disfavors a high volatile solubility in silicate magma, and degassing would therefore be expected to be nearly complete in the earliest stages of melting and segregation \citep{Monteux2018}.

Finally, the melt in our \emph{magma ocean} and \emph{magma sill} models cannot breach the cold surface layers, as the temperature is too low for melt to exist in the porous medium. Our simulations cannot resolve potential melt transport through the upper lid by fracturing \citep{2013GeoJI.195.1406K} or gravitational instability in the layered structure in Figure \ref{fig:cbulk}, which may bury the primitive lid \citep{2012ChEG...72..289W}, and lead to efficient heat loss and magmatic resurfacing. This would decrease the total retention of magma on the inside of planetesimals. However, this does not affect our conclusions or any constraints on temperature inversions of planetesimals unless this transport is faster or comparable to the magma segregation in the interior.

\section{Summary \& conclusions}
\label{sec:conclusions}

In this study we investigated magma genesis and redistribution in planetesimals during and shortly after the protoplanetary disk phase. Using scaling relations we demonstrated that the interior evolution of planetesimals sensitively depends on a variety of model parameters, with the grain size exerting the primary control on melt segregation. Based on average chondritic abundances of the most common mineral groups in meteorites, we calculated the composition for rock--melt aggregates comprising idealized components using a reactive, multi-component melting model. We quantified the effects on $^{26}$Al partitioning and magma ascent with a coupled, two-phase flow model. We defined the melt segregation number $\mathrm{R}_{\mathrm{seg}}$ as the ratio between the heating and melt transport time scales in a planetesimal, which establishes the leading order parametric control on the propensity for magma redistribution during the heating stage of planetesimal evolution. We predicted that the primary two controls are the melt--rock density contrast $\Delta \rho$ and the mineral grain size $a_{0}$.

Investigating the relative importance of model parameters for the evolution of planetesimals, we categorize model outcomes in terms of their melt segregation numbers $\mathrm{R}_{\mathrm{seg}}$ and their formation times $t_{\mathrm{form}}$. Using this scheme, we find:

\begin{itemize}
	\item Planetesimal melt migration behavior can be classified in three general regimes:
	\begin{itemize}
		\item The \emph{magma ocean} regime with global interior magma oceans, where rapid melting overwhelms melt segregation.
		\item The \emph{magma sill} regime, where global interior magma oceans are prevented by rapid magma transport that concentrates melt in sills beneath the cool lid.
		\item The \emph{undifferentiated} regime with a low degree of melting, minor melt segregation, and therefore chemically primordial and largely undifferentiated interiors.
	\end{itemize}
	\item \emph{Magma sill} models crystallize to a compositionally stratified structure, with shallow depth layers dominantly enriched in the low melting point components, feldspar and pyroxene, and a paucity of high melting point components, such as olivine \citep[cf.][]{1998Icar..134..187G,mizzon2015magmatic}. The crystallization sequence, and thus the compositional stratification, however, may be affected by convective motions beyond the disaggregation limit, which we do not model here.
	\item \emph{Magma ocean} and \emph{magma sill} models show temperature inversions for high $\mathrm{R}_{\mathrm{seg}}$ and early $t_{\mathrm{form}}$, where the temperatures in the shallow- to mid-mantle are higher than at the center of the planetesimal. These thermal inversions, however, are restricted to formation times $t_{\mathrm{form}}$ $\lesssim$ 1 Myr and $\mathrm{R}_{\mathrm{seg}}$ $\gtrsim 1.5$ for $\Delta T \geq 250$ K. Therefore, the majority of planetesimals likely underwent thermal evolutionary scenarios that can be qualitatively reproduced with models that do not take into account melt segregation and $^{26}$Al partitioning, depending on the level of detail that needs to be assessed.
	\item The \emph{magma sill} regime can be achieved depending on a combination of a few key parameters:
	\begin{itemize}
		\item The formation time $t_{\mathrm{form}}$ controls the total amount of energy available from $^{26}$Al and restricts the regime for melt segregation to 0.5 $\times$ $t_{1/2,^{26}\mathrm{Al}}$ $\lesssim$ $t_{\mathrm{form}}$ $\lesssim$ 1.75 $\times$ $t_{1/2,^{26}\mathrm{Al}}$, with a peak at $\approx$ 1 Myr after CAIs, when the rate of melt transport dominates over the generation of new melts.
		\item The grain size $a_{0}$ controls the rate of melt transport and thus whether a planetesimal with sufficient heating evolves towards a melt segregated structure. Below $a_{0}$ $\lesssim$ 0.1 mm no segregation occurs; above $a_{0}$ $\gtrsim$ 1 mm segregated structures form.
		\item The solid--melt density contrast $\Delta \rho$ is of secondary importance, but can enhance melt segregation in the regime transition from $a_{0}$ = 0.1 mm to 1 mm.
	\end{itemize}
\end{itemize}

To conclude, in this paper we have advanced the technical capabilities to simulate multi-phase and multi-component planetesimal evolution, gaining insights into features not accessible to single-phase fluid dynamics models. However, unraveling more detailed evolutionary regimes of planetesimals will require a time-dependent treatment that includes metal and volatile phases, which shape the structure and subsequent evolution of these bodies \citep{2018arXiv180900079K}. Further work is required to understand planetesimal evolution and its connection to the meteoritic record and rocky planet formation \citep{2018Icar..302...27L}. In particular, robust scaling laws for the evolution of grain sizes in partially molten and heated systems relevant for planetesimal settings are required to establish more detailed regimes for melt segregation.

In the mid-term, future spacecraft missions \citep{2017NatAs...1E..95A} may be able to deliver further observational constraints on asteroid-belt objects. In conjunction with self-consistent evolutionary models of metal--silicate and solid--melt segregation, these can help to further decipher the interior evolution of planetary bodies in the early solar system and sharpen our understanding of terrestrial planet formation in the solar system and elsewhere.

\paragraph{Acknowledgements}
The authors acknowledge stimulating discussions with Ian Sanders and Guy Consolmagno, and constructive comments by David Bercovici and an anonymous referee, which helped to improve the manuscript. \textsc{TL} was supported by ETH Z{\"u}rich Research Grant ETH-17 13-1 and acknowledges partial financial support from a MERAC travel grant of the Swiss Society for Astrophysics and Astronomy and from the National Center for Competence in Research PlanetS supported by the Swiss National Science Foundation. \textsc{TK} acknowledges financial support from Jenny Suckale, Stanford University. \textsc{TK} and \textsc{RFK} received funding from the European Research Council under the European Union's Seventh Framework Programme (FP7/2007--2013)/ERC grant agreement number 279925.  The numerical simulations in this work were performed on the \textsc{EULER} computing cluster of ETH Z{\"u}rich, and were analyzed using the open source software environments \textsc{matplotlib}\footnote{\href{https://matplotlib.org}{matplotlib.org} \citep{matplotlib}} and \textsc{seaborn}\footnote{\href{https://seaborn.pydata.org}{seaborn.pydata.org}}.

\section*{Supplementary Materials}
\label{sec:suppl}

A simulation video associated with this article can be found attached to the \href{https://doi.org/10.1016/j.epsl.2018.11.034}{publication} and at \href{https://vimeo.com/243441022}{this URL}. The video shows a comparison between \emph{magma ocean} and \emph{magma sill} end-member models. Magma ocean stages are indicated in yellow. The {\it H3} annotation describes the heating value below which radioactive heating from $^{26}$Al is inefficient. Shown are various parameters for both models, from left to right: melt fraction $\phi$, temperature $T$, radiogenic heating $H$, melt/liquid upwelling velocity $w_\mathrm{liq}$, composition fraction of liquid $c_\mathrm{liq}^\mathrm{i}$, and composition fraction of solid $c_\mathrm{sol}^\mathrm{i}$. The composition is broken-down into the defined pseudo-components {\em fsp}, {\em pxn}, and {\em olv}. The model is started (= planetesimal formation time) at 0.9 Myr after CAIs.



\end{document}